# Towards Developing Network forensic mechanism for Botnet Activities in the IoT based on Machine Learning Techniques

Nickilaos Koroniotis[1], Nour Moustafa[1], Elena Sitnikova[1], Jill Slay[1]

[1]School of Engineering and Information Technology
University of New South Wales Canberra, Australia
n.koroniotis@student.adfa.edu.au
nour.moustafa@unsw.edu.au
e.sitnikova@adfa.edu.au
j.slay@adfa.edu.au

**Abstract.** The IoT is a network of interconnected everyday objects called "things" that have been augmented with a small measure of computing capabilities. Lately, the IoT has been affected by a variety of different botnet activities. As botnets have been the cause of serious security risks and financial damage over the years, existing Network forensic techniques cannot identify and track current sophisticated methods of botnets. This is because commercial tools mainly depend on signature-based approaches that cannot discover new forms of botnet. In literature, several studies have conducted the use of Machine Learning (ML) techniques in order to train and validate a model for defining such attacks, but they still produce high false alarm rates with the challenge of investigating the tracks of botnets. This paper investigates the role of ML techniques for developing a Network forensic mechanism based on network flow identifiers that can track suspicious activities of botnets. The experimental results using the UNSW-NB15 dataset revealed that ML techniques with flow identifiers can effectively and efficiently detect botnets' attacks and their tracks.

## 1 Introduction

An increasingly popular new term, is the Internet of Things (IoT). The concept of IoT dates back to the early 1980s, where a vending machine selling Coca-Cola beverages located at the Carnegie Mellon University was connected to the Internet, so that its inventory could be accessed online to determine if drinks were available [33]. Today, the IoT is an umbrella term, covering a multitude of devices and technologies, that have both Internet capabilities, and serve some primary function, such as: home automation, including smart air conditioning system, smart fridge, smart oven and smart lamps, wearable devices (i.e., smart watch and fitness tracker), routers, healthcare, DVRs, smart cars, etc. In general, IoT can be viewed as a collection of devices with low processing power and some form of network communication capabilities that allow a user to remotely access and use their services or view information collected by them [33][34].

Recently, a number of malware have appeared that target IoT, as hackers and researchers alike have noticed that in the majority, IoT devices are mostly vulnerable to the simplest attacks, as displayed by Mirai, a botnet consisting of 100.000 infected "things" that in October 2016, attacked and took out a good portion of the Internet's high-profile services such as Twitter and Netflix by doing a DDoS attack on Dyn (DNS provider) [36]. Since then, Mirai has slowly been divided into smaller botnets, and new botnets have risen, such as BrickerBot, which as its name implies "bricks" an IoT device (permanently disables it) and Hajime, a vigilante botnet which has been observed to infect devices targeted by Mirai, and "securing" the device, not allowing another malware to infect it. Clear methods have to be developed, in order to effectively mitigate and investigate at a forensic level IoT devices, as it is apparent that IoT is ever increasing in popularity, both by consumers (including companies) and hackers alike [35].

It is commonly known that the Internet is not a safe place, being full of security threats, with consequences ranging from mere inconvenience, to possible life-threatening scenarios. Amongst these threats, a family of cyber threats called Botnets, considered to be the one of the most destructive capabilities [21]. A botnet is defined as a set of internet-based appliances, which involves computer systems, servers and Internet of Thing (IoT) devices infected and managed by a common type of attacks, such as Denial of Service (DoS), DDoS, phishing attacks [16] [21]. Bots differ from other malware, in that they include a channel of communication with their creators, allowing them to issue commands to their network of bots (i.e., Zombies) and thus making botnets versatile when it comes to their functionality [1][2].

Lately, there have been some prominent examples of botnets that harnessed the aggregated processing power of the IoT, and many infection routs have been revealed. One example is through the Internet, with portions of a botnet actively scanning the Internet for vulnerable devices, gaining access to them and then allowing a third device to infect the victim with the botnet malware [16]. Another is by using close proximity networks to gain direct communication with a device, infect it and allow the malware to propagate to the rest of the IoT devices in the immediate vicinity [17].

Different security controls have been used for defining botnet events, including Network forensic techniques and tools and intrusion detection and prevention systems. The existing techniques and tools basically use the principle of expert system which is generating specific rules in a blacklist and matching the upcoming network traffic against those rules. In this study, we investigate and analyze the role of ML techniques to construct a Network forensic technique based on network flow identifiers (i.e., source and destination IP address/ports and protocols). According to [22] [23], Network forensic is a branch of the digital forensics in order to monitor and inspect network traffic for defining the sources of security policy abuses and violations.

Machine learning techniques, learn and validate given network data for classifying legitimate and anomalous observations, have been utilized for building Network forensic techniques, but there are still two challenges should be addressed: producing high false alarm rates and defining the paths of attacks, in particular botnet events [6] [22] [23] [24] [25]. Machine learning algorithms include clustering, classification, pattern recognition, correlation, statistical techniques [6] [24] [26]. In clustering techniques, network traffic is separated into groups, based on the similarity of the data, without the need to pre-define these groups, while classification mechanisms learn and test patterns of network data associated with their class label [6].

The main contribution of this paper is the use of four classification techniques, so-called, Decision Tree C4.5 (DT), Association Rule Mining (ARM), Artificial Neural Network (ANN) and Naïve Bayes (NB) for defining and investigating the origins of botnets. The four techniques are used to recognize attack vectors, while the origins of attacks are linked with their flow identifiers as a Network forensic mechanism.

The rest of the paper is organized as follows. Section 2 discusses the background and related work then section 3 explains the Network forensic architecture and its components. In section 4, the experimental results are displayed and discussed and finally, the conclusion is given in section 5.

## 2 Background and previous studies

This section discusses the background and related studies for the IoT, Botnets and Network forensics.

### 2.1 IoT

Even though IoT is slowly been assimilated in one form or another in everyday life, there is no doubt that there exist a number of issues with respect to security and privacy. One such example, is displayed by Eyal Ronen et al. [37]. The researchers investigate a possible attack vector for Philips Hue smart lamps, and a way that these IoT devices, which primarily use Zigbee as a communication protocol for the lamps to communicate with each other and their controllers. The attack vector they proposed and tested, takes advantage of exploits found in such devices and proposed that, under certain conditions (number of devices and relative distance), a malware can spread through a city, "jumping" from device to device, or even permanently disable ("brick") them.

A more general point of view was adopted by M Mahmud Hossain, et al. [38], who provided an analysis on open security problems in the IoT. They showed, through the literature, that IoT devices have been proven to be insecure, with researchers successfully compromising them with relative ease. Then, based on the IoT ecosystem, which is made up of: IoT devices, Coordinator, Sensor Bridge Device, IoT Service, Controller, they proceeded to highlight that conventional security and forensics mechanisms are inapplicable in IoT, as such devices are constrained in a number of ways (processing power, battery life, Network mobility), as-well-as diverse (IoT devices range from simple sensors, to complex Personal Computers) and research on those areas should become a priority. On the other hand, Yin Minn Pa Pa, et al. [39], made observations on the nature and type of IoT devices being actively scanned through the Internet and went on to propose an IoT specific Honeypot named IoTPOT. Their solution included an IoTBOX a collection of virtual IoT machines (Linux OS), which helps make the Honeypot appear as a legitimate device. They observed a number of malware in action, most interestingly, they observed a repeating pattern, where a single host would perform the intrusion and information gathering process and the actual infection would be handled by a different host. Through the literature, it is evident that the field of IoT still being developed and as such various issues with this new and growing field are being discovered daily.

**2.2 Botnets in IoT**

In the literature, a variety of techniques that researchers utilize to understand Botnets and to study them has been observed. In their work, though not specifically related to IoT Botnets, Ashkan Rahimian, et al. [40] studied a Bot named Citadel. To specify, they employed several code analysis techniques to measure the similarities between Zeus and Citadel (Citadel being a "descendant" of the Zeus Bot) and in order to make the reverse engineering process faster they proposed a new approach named clone-based analysis, which attempts to identify parts of the malware's code that originated from a different source, thus reducing the amount of code an analyst needs to review. Botnet scanning and identification was the main focus of Amir Houmansadr, et al. [41], who introduced BotMosaic, a tool following an architecture similar to Client-Server. Among other things BotMosaic, performs non-distorting network flow watermarking, for detecting IRC Botnets, which is the non-altering (towards network traffic content) process of "Marking" traffic so that it can be identified at a later time period. Although many techniques have been developed, in order to scan, fingerprint, identify and generally investigate a Botnet, as time moves forward, malware authors adapt and make their Bots more difficult to identify, this combined with the unique nature of the IoT, produces issues which need to be addressed.

Botnets are capable of launching a number of attacks, like Distributed Denial of Service attacks (DDoS), Keylogging, Phishing and Spamming, Identity theft and even other Bot proliferation [4]. Understandably, some research has been conducted in developing ways to detect and identify the presence of botnets in a network. One such way is the utilization of machine learning techniques on captured packets (Stored in files like pcap files) which are often grouped into network flows (Netflows), in an attempt to distinguish between legitimate user traffic and botnet traffic [5].

A botnet's malware gets delivered to vulnerable targets through what is known as a propagation mechanism. Most commonly there exist two types of propagation, passive and active. Passive or self-propagation techniques rely on the bots themselves to actively scan the Internet for signs of vulnerable devices and then attempt to exploit the identified vulnerability, turning the vulnerable hosts into bots themselves [2][3]. Passive propagation techniques, require users to access social media posts, storage media or websites that have been compromised and through user interaction, such as accessing URLs or other active parts of a website, download the malware (bot) to their machine, infecting it and making it part of the botnet [2][3].

Various studies [18][19][22][23][24][25][26] have employed Machine learning techniques, to distinguish between normal and botnet network traffic and designing Network forensic techniques and tools. In their work, Roux et al [18], created an intrusion detection system for IoT which takes into account wireless transmission data through probes. The information is collected, which is relevant to signal strength and direction, determining that a signal originating from an unexpected direction, such as outside of the location that is being monitored, is deemed illegitimate. The classification of whether the observed signal indicates an attack or not is produced by a neural network. In another approach, Lin et al [19], employed Artificial Fish Swarm Algorithm to produce the optimal feature set, which was then provided to a Support Vector Machine which detected botnet traffic. They reported a slight increase in accuracy when compared with Genetic Algorithms for feature selection, but produced great improvement time-wise. On the other hand, Greensmith et al [20], proposed the utilization of Artificial Immune Systems as a way of securing the IoT. They propose a combination of multiple AIS, in order to handle the heterogeneity of the IoT.

**2.3 Network forensics in IoT**

Network Forensics problem is governed by low availability of traces and evidence. For instance, Rana Khattak, et al. [27] tackled the problem of the ever-increasing size of network log files, by using parallel execution through Hadoop's MapReduce. S.Bansal et al. [28] proposed their own generic framework for detecting Botnets, focusing on Network forensics techniques, such as packet capturing and inspection, which has the structure of a balanced framework, although it appears to be quite theoretical in nature.

Saied et al [29] employed an Artificial Neural Network for developing a distributed topology of DDoS inspectors residing in different networks, to detect DDoS attacks, based on timing and header inspection. Divakaran et al [30] developed their own Framework for detecting such attacks (DDoS), by employing a regression model based on defined patterns. They did so, by grouping packets into network flows, and flows into sessions, based on timing of packet arrival, and then through regression, they identify anomalous patterns in traffic, also providing information on detection rates for different botnets.

On a different note all together, Wang et al [31] proposed an attack detection and forensics technique to tackle the threat introduced by malware. In their approach, an attack detection model locates the presence of an incident after which, it cooperates with a honeypot to gather relevant information, which are then used by a forensics module to carry out an automated forensics procedure. Their work is interesting, as they effectively make the forensic process more active than what it usually is. As the Internet is incorporated in an ever-growing number of technologies that are gradually adopted by society, it should come as no surprise that Network forensics adopts a more integral role in investigating malicious activity.

## 3. Network forensic architecture and components

The proposed Network forensics mechanism includes four components: 3.1) traffic collection, 3.2) network feature selection, 3.3) machine learning techniques, and 3.4) evaluation metrics, as depicted in Fig. 1 and explained below.

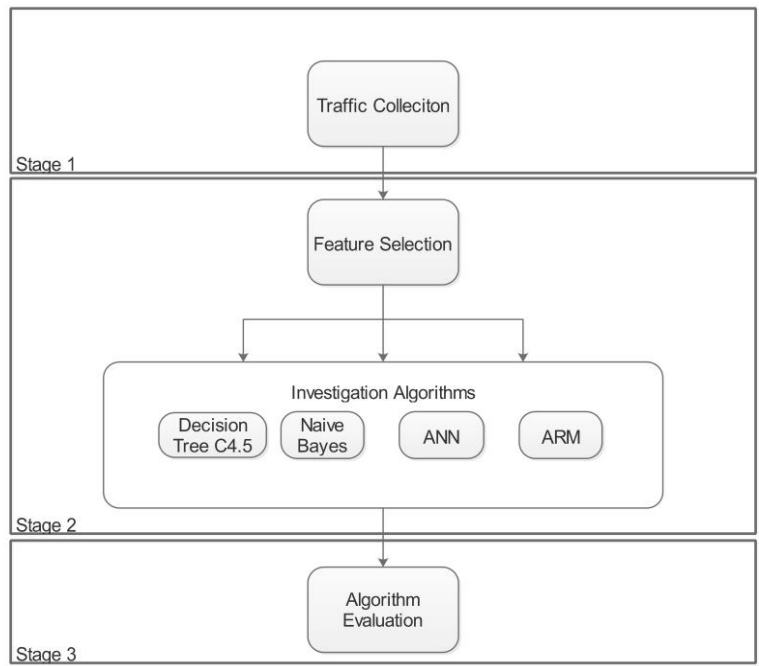

Fig. 1: Proposed Network forensic architecture

**3.1 Traffic collection**

The immense volume of network traffic generated by todays networks, makes it necessary for a way to aggregate and summarize the captured packets, allowing for easier storage so that they can be used in construction of Network forensic mechanisms. Raw network packets are captured by using a tcpdump tool, which can access the network interface card to do so [15]. The next step is feature generation. By using tools like Bro and Argus, features are generated from the raw packets that were previously captured. The features in the UNSW-NB15 dataset were generated in this manner [15].

The network sniffing process needs to be conducted at key points of the network, particularly, at ingress routers, to make sure that the network flows that are gathered are relevant, which is determined by the source/destination IP address and protocol.

This process helps, in investigating the origins of cyber-related incidents, and lowers the necessary processing time.

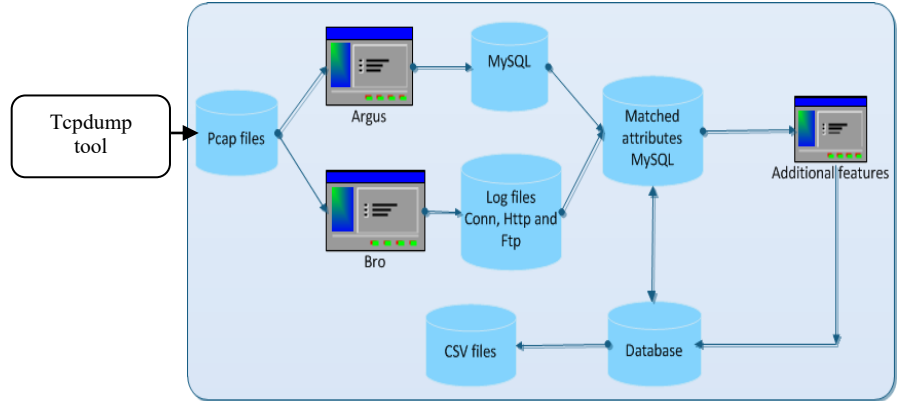

Fig.2: collecting network features of UNSW-NB15 dataset [15]

### 3.2 Network feature selection method

Feature selection is the method of adopting important/relevant attributes in a given dataset. Method of feature selection is classified into filter, wrapper and hybrid of the first two. A filter feature selection mechanism denotes selecting suitable features without the use of class label, while a wrapper one depends on ML techniques [8][9]. An example of filtering methods for feature selection, is Information Gain (IG) and Chi-square ($x^2$). Wrapper methods study the performance of an algorithm that will be used in the end, as a criterion for selecting a suitable subset of the existing features. Intuitively, these methods split the features into subsets, and use these subsets to train the model that will be used after the pre-processing stage, using the error rate to score the subsets. Hybrid methods combine filter and wrapper methods to perform feature selection during the execution of machine learning algorithms.

We use the information gain feature mechanism as it is one of the simplest methods that can adopt relevant features in large-scale data, as in network datasets. More precisely, Information Gain (IG) selects features, by calculates the apparent reduction in entropy, when the feature to be selected is used to split the dataset.

### 3.3 Machine learning techniques
For the classification stage, we use the Weka tool [42] for applying four well-known machine learning algorithms. These algorithms are briefly described as follows.

- **Association Rule Mining (ARM)** [14]- is a classification algorithm, which is performed by generating rules of a form similar to $\{V_1,V_2,…,V_n\} => \{C_1\}$, where $V_{1-n}$ are values of features and $C_1$ is a class value.

- **Artificial Neural Network (ANN)** [11][12]- is a classification model which was based on the idea of the human neurons [11][12]. They usually are comprised of a number of neurons, which have multiple input flows and a single output, with some variants having multiple layers of units. The simplest form of an ANN is called a perceptron, taking the vector of attributes as input and classifying it.

- **Naïve Bayes (NB)** [13]- classifies a record $R_1$ (collection of features) into a specific class $C_2$, if and only if the probability of that record to belong to that specific class, with respect to the record is greater than the probability of the record belonging to another class That is, $P(C_2/R_1) > P(C_n/R_1)$, with $C_n$ being any class other than $C_2$.

- **Decision Tree C4.5 (DT)** [11]- is a classification algorithm which produces a tree-like structure to determine the class chosen for a record. The attributes are used as the nodes of the tree and criteria are formulated leading from one node to the next, with the leaves being the Class value that is assigned to the record [11].

### 3.4 Evaluation metrics

We utilize the confusion matrix [32] as a way of comparing the performance of the ML algorithms presented in the previews section. An example of a confusion matrix is given in Table 1. In simple terms, it is a table which depicts the possible outcomes of a classification, which in our case is either '1', there was an attack detected or '0' normal network traffic, against the actual values of the class feature already present in the evaluation (testing) dataset.

There are four condition that can be shown in a confusion matrix, True Positive (TP), where the classifier has correctly identified the class feature and the value of that feature is positive (in our case there was an attack detected), True Negative (TN), similar to TP but the value of the class feature is negative (normal traffic), False Positive (FP), where the classifier identifies a record as an attack when, in actuality it is normal traffic and False Negative (FN), which incorrectly classifies an attack record as normal traffic.

Table 1: Confusion matrix

|  | Actual Negative | Actual Positive |
|---|---|---|
| Predicted Negative | TN | FP |
| Predicted Positive | FN | TP |

By combining the TP, TN, FP, FN values we are able to create two metrics, namely **Accuracy** and **False Alarm Rate**, which we can use to evaluate the Classifiers. These two metrics are calculated as follows:

- **Accuracy** represents the probability that a record is correctly identified, either as attack, or as normal traffic. The calculation of Accuracy (Overall Success Rate) is OSR= (TN+TP)/(TP+FP+TN+FN)

- **False Alarm Rate (FAR)** represents the probability that a record gets incorrectly classified. The calculation of the False Alarm Rate is FAR = FP+FN/(FP+FN+TP+TN)

## 4. Experimental results and discussions

### 4.1 Dataset used for evaluation and feature selection

In order to compare the four aforementioned algorithms, we used the UNSW-NB15 dataset was designed at the Cyber Range Lab of the Australian Center of Cyber Security at UNSW Canberra [15]. We selected UNBS-NB 15, as it is one of the newest datasets in wide use today, thus providing accurate representations of both conventional (not malicious) network traffic, as-well-as a number of network attacks performed by Botnets. The dataset was produced by making use of the IXIA PerfectStorm tool, which produced a mixture of legitimate user network traffic and attack traffic, with the latter being categorized into 9 groups, Fuzzers, Analysis, Backdoor, DoS, Exploits, Generic, Reconnaissance, Shellcode, Worms.

Table 2: UNSW-NB15 Features selected with Information Gain

| Ranking | Feature selected | Feature description |
|---------|------------------|---------------------|
| 0.654 | sbytes | Source to destination transaction bytes |
| 0.491 | dbytes | Destination to source transaction bytes |
| 0.477 | smean | Mean packet size transmitted by source |
| 0.464 | sload | Source bits per second |
| 0.454 | ct_state_ttl | |
| 0.444 | sttl | Source to destination time to live value |
| 0.439 | dttl | Destination to source time to live value |
| 0.429 | rate | |
| 0.409 | dur | Record total duration |
| 0.406 | dmean | Mean packet size transmitted by destination |

A short description of these attacks is given here [43]. Fuzzers: where an attacker attempts to identify security weaknesses in a system, by providing large quantities of randomized data, expecting it to crash. Analysis, comprised of a variety of intrusion techniques targeting ports, email addresses and web scripts. Backdoor: a method of bypassing authentication mechanisms, allowing unauthorized remote access to a device. DoS: a disruption technique, which attempts to bring the target system in a state of non-responsiveness. Exploit: a combination of instructions, that take advantage of bugs in the code, leading the targeted system/network in a vulnerable state. Generic: an attack that attempts to cause a collision in a block-cipher using a hash function. Reconnaissance: an information gathering probe, usually launched before the actual attack. Shellcode: an attack during which carefully crafted commands are injected in a running application, through the net, allowing for further escalation by taking control of the remote machine. Worm: an attack based on a malware that replicates itself, thus spreading itself in a single or multiple host. The dataset is comprised of 49 features, including the class feature, and the portion of it that we will be making use, contains 257,673 (created by combining the training and testing datasets).

To test the classifiers, we performed Information Gain Ranking Filter (IG) for selecting the highest ten ranked features as listed in Table 2.

### 4.2 Performance evaluation of ML algorithms

The confusion matrices of the four classification algorithms are listed in Tables 3-6 on the training and testing sets of the UNSW-NB15 dataset. The Weka tool was used for applying the four techniques using the default parameters with a 10-fold cross validation in order to effectively measure their performance.

Our experiments show that Decision Tree C4.5 Classifier was the best at distinguishing between Botnet and normal network traffic. This algorithm makes use of Information Gain, to pick the feature which best splits the data based on the classification feature, during construction of the tree and at every node. The test showed that DT had the highest accuracy out of all the algorithms that were tested at 93.23%, and the lowest FAR at 6.77%.

ARM was the second-best classifier, having an accuracy of close to 86% and FAR just over twice that of the DT. The Naïve Bayes classifier, which relies on probability to classify records in classes was third, with 20% less accuracy and close to 21% more false alarms than the DT. Finally, the Artificial Neural Network was the least accurate out of the four algorithms that we tested, with accuracy and false alarm rate for this classifier showing a 30% differentiation from the C4,5 algorithm.

Table 3: ARM confusion matrix

| Normal | Attack | Prediction/Actual |
|--------|--------|-------------------|
| 31785  | 10894  | Normal            |
| 12675  | 108654 | Attack            |

Table 4: DT confusion matrix

| Normal | Attack | Prediction/Actual |
|--------|--------|-------------------|
| 84607  | 8393   | Normal            |
| 9058   | 155615 | Attack            |

Table 5: NB confusion matrix

| Normal | Attack | Prediction/Actual |
|--------|--------|-------------------|
| 84101  | 8899   | Normal            |
| 61380  | 103293 | Attack            |

Table 6: ANN confusion matrix

| Normal | Attack | Prediction/Actual |
|--------|--------|-------------------|
| 2719   | 90281  | Normal            |
| 2562   | 162111 | Attack            |

Table 7: performance evaluation of four techniques

| **Classifier** | **Accuracy** | **FAR** |
|----------------|--------------|---------|
| ARM            | 86.45%       | 13.55%  |
| DT             | 93.23%       | 6.77%   |
| NB             | 72.73%       | 27.27%  |
| ANN            | 63.97%       | 36.03%  |

A similar comparison of Machine Learning performance was conducted on the KDD99 dataset, arguably one of the most popular datasets still in use for security related work, like evaluating Network Intrusion Detection Systems [43].

Table 8: ANN confusion matrix

| **Classifier** | **Accuracy** | **FAR** |
|----------------|--------------|---------|
| ARM            | 92.75%       | -       |
| DT             | 92.3%        | 11.71%  |
| NB             | 95%          | 5%      |
| ANN            | 97.04%       | 1.48%   |

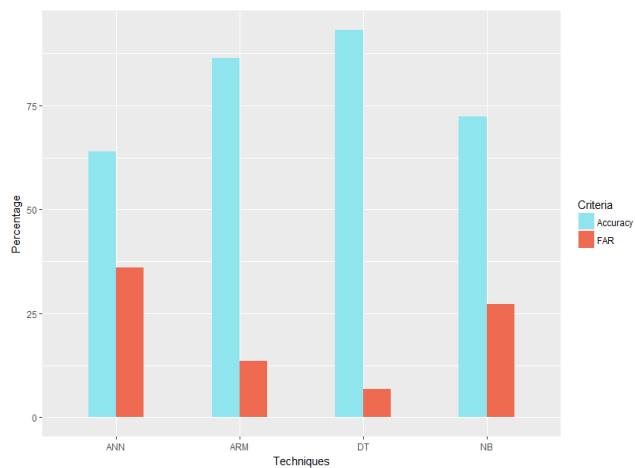

Fig.3. accuracy vs. FAR of four classifiers

By combining the identifiers of a network flow, with the corresponding condition Label, which depicts the result of the classification techniques mentioned previously that classify a record under "attack" or "normal", the tracking of attack instances becomes possible. An example of the final form of the dataset is given in Table 8, which provides a number of flows and their classification label, taken from the UNSW-NB15 dataset. The network forensic technique that this paper illustrates, can assist network administrators, security experts or even law enforcement, to identify, track, report and even mitigate security incidents that threaten the security of their network.

To produce the results depicted in Table 8, first the combination of flow information and Association Rule Mining needs to be performed, as shown in figure 8. The produced rules indicate the type of attack and the means by which it was identified (features used in the rule), and later by combining that information with a row containing source/destination IP address/ Port numbers and protocol used, it becomes possible to attribute specific botnet activities to specific hosts. Such association between flow information and rule, can be of vital importance in monitoring botnet activities, through identifying infected hosts and showing their actions over time.

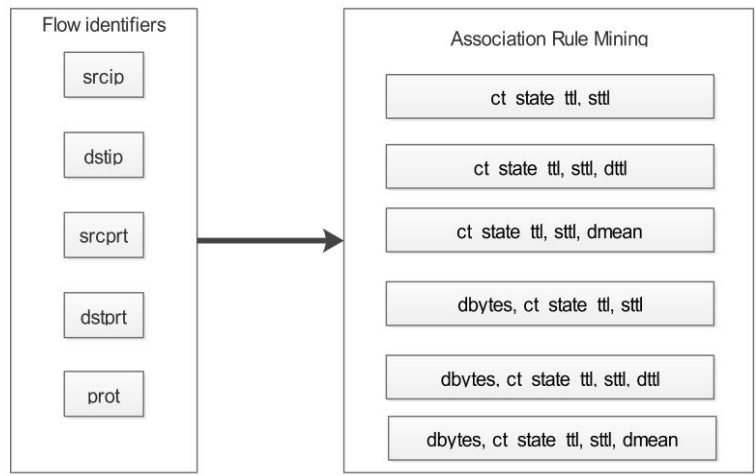

Fig 4. Combining Flow data with Rules.

Table 9: flow identifiers associated with actual label for investigating attacks

| srcip | sport | dstip | dsport | proto | Label |
|---|---|---|---|---|---|
| 149.171.126.14 | 179 | 175.45.176.3 | 33159 | tcp | 0 |
| 149.171.126.18 | 1043 | 175.45.176.3 | 53 | udp | 0 |
| 175.45.176.3 | 46577 | 149.171.126.18 | 25 | tcp | 1 |
| 149.171.126.15 | 1043 | 175.45.176.3 | 53 | udp | 0 |
| 175.45.176.2 | 16415 | 149.171.126.16 | 445 | tcp | 1 |

## 5. Conclusions

This paper discusses the role of machine learning techniques for identifying and investigating botnets. There are four ML techniques of DT, ANN, NB and ANN machine are evaluated on the USNW-NB15 dataset. The accuracy and false alarm rate of the techniques are assessed, and the results revealed the superiority of the DT compared with the others. The best machine learning techniques and flow identifiers of source/destination IP addresses and protocols can effectively and efficiently detect botnets and their origins as Network forensic mechanism.

## Acknowledgements

Nickolaos Koroniotis would like to thank the Commonwealth's support, which is provided to the aforementioned researcher in the form of an Australian Government Research Training Program Scholarship.

## References

1. Silva, S. S. C.; Silva, R. M. P.; Pinto, R. C. G. & Salles, R. M. (2013), 'Botnets: A survey', Computer Networks **57**(2), 378 - 403.
2. Khattak, S.; Ramay, N. R.; Khan, K. R.; Syed, A. A. & Khayam, S. A. (2014), 'A Taxonomy of Botnet Behavior, Detection, and Defense', IEEE Communications Surveys Tutorials **16**(2), 898-924.
3. Negash, N. & Che, X. (2015), 'An Overview of Modern Botnets', Information Security Journal: A Global Perspective **24**(4-6), 127-132.
4. Amini, P.; Araghizadeh, M. A. & Azmi, R. (2015), A survey on Botnet: Classification, detection and defense, in 'Electronics Symposium (IES), 2015 International', pp. 233--238.
5. Goodman, Nathan. (2017) "A Survey of Advances in Botnet Technologies." arXiv preprint arXiv:1702.01132.
6. Amini, Pedram, Muhammad Amin Araghizadeh, and Reza Azmi (2015) "A survey on Botnet: Classification, detection and defense." Electronics Symposium (IES), 2015 International. IEEE.
7. Sheen, Shina, and R. Rajesh 2008 "Network intrusion detection using feature selection and Decision tree classifier." TENCON 2008-2008 IEEE Region 10 Conference. IEEE.
8. Chandrashekar, Girish, and Ferat Sahin. (2014) "A survey on feature selection methods." Computers & Electrical Engineering 40.1: 16-28.
9. Jović, Alan, Karla Brkić, and Nikola Bogunović (2015) "A review of feature selection methods with applications." Information and Communication Technology, Electronics and Microelectronics (MIPRO), 2015 38th International Convention on. IEEE.

10. Bhavsar, Yogita B., and Kalyani C. Waghmare (2013) "Intrusion detection system using data mining technique: Support vector machine." International Journal of Emerging Technology and Advanced Engineering 3.3: 581-586.
11. Area, Sonepat, and Ranchi Mesra (2012) "Analysis of Bayes, Neural Network and Tree Classifier of Classification Technique in Data Mining using WEKA.".
12. Sebastian, Sumam, and Jiby J. Puthiyidam (2015) "Evaluating students performance by artificial neural network using weka." International Journal of Computer Applications 119.23.
13. Xiao, L., Chen, Y. and Chang, C.K. (2014) "Bayesian model averaging of Bayesian network classifiers for intrusion detection." In Computer Software and Applications Conference Workshops (COMPSACW), 2014 IEEE 38th International (pp. 128-133). IEEE.
14. Moustafa, Nour, and Jill Slay. "The significant features of the UNSW-NB15 and the KDD99 data sets for Network Intrusion Detection Systems." Building Analysis Datasets and Gathering Experience Returns for Security (BADGERS), 2015 4th International Workshop on. IEEE, 2015.
15. Moustafa, Nour, and Jill Slay (2015) "UNSW-NB15: a comprehensive data set for network intrusion detection systems (UNSW-NB15 network data set)."Military Communications and Information Systems Conference (MilCIS), 2015. IEEE.
16. Pa, Y.M.P., Suzuki, S., Yoshioka, K., Matsumoto, T., Kasama, T. and Rossow, C., 2015. "IoTPOT: analysing the rise of IoT compromises." EMU, 9, p.1.
17. Ronen, E.; Shamir, A.; Weingarten, A. O. & O'Flynn, C. (2017), IoT Goes Nuclear: Creating a ZigBee Chain Reaction, in '2017 IEEE Symposium on Security and Privacy (SP)', pp. 195-212.
18. Roux, J., Alata, E., Auriol, G., Nicomette, V. and Kaâniche, M., (2017) "Toward an Intrusion Detection Approach for IoT based on Radio Communications Profiling." In 13th European Dependable Computing Conference.
19. Lin, K.C., Chen, S.Y. and Hung, J.C., (2014). "Botnet detection using support vector machines with artificial fish swarm algorithm." Journal of Applied Mathematics, 2014.
20. Greensmith, J (2015) "Securing the Internet of Things with responsive artificial immune systems." In Proceedings of the 2015 Annual Conference on Genetic and Evolutionary Computation (pp. 113-120). ACM.
21. Pijpker, Jeroen, and Harald Vranken (2016) "The role of Internet Service Providers in botnet mitigation." Intelligence and Security Informatics Conference (EISIC), 2016 European. IEEE.
22. Wang, Xiao-Jing, and Xiao-yin Wang (2010) "Topology-assisted deterministic packet marking for IP traceback." The Journal of China Universities of Posts and Telecommunications 17.2: 116-121.
23. Khan, S., Gani, A., Wahab, A.W.A., Shiraz, M. and Ahmad, I. (2016) "Network forensics: review, taxonomy, and open challenges." Journal of Network and Computer Applications, 66, pp.214-235.
24. Moustafa, Nour, Jill Slay, and Gideon Creech (2017) "Novel Geometric Area Analysis Technique for Anomaly Detection using Trapezoidal Area Estimation on Large-Scale Networks." IEEE Transactions on Big Data.
25. Prakash, P. Banu, and ES Phalguna Krishna (2016) "Achieving High Accuracy in an Attack-Path Reconstruction in Marking on Demand Scheme." i-Manager's Journal on Information Technology 5.3: 24.
26. Sangkatsanee, Phurivit, Naruemon Wattanapongsakorn,,Chalermpol Charnsripinyo (2011) "Practical real-time intrusion detection using machine learning approaches." Computer Communications 34.18: 2227-2235.

27. Moustafa, Nour, Gideon Creech, and Jill Slay. "Big Data Analytics for Intrusion Detection System: Statistical Decision-Making Using Finite Dirichlet Mixture Models." Data Analytics and Decision Support for Cybersecurity. Springer, Cham, 2017. 127-156.
28. S. Bansal, M. Qaiser, S. Khatri and A. Bijalwan (2015) "Botnet Forensics Framework: Is Your System a Bot," 2015 Second International Conference on Advances in Computing and Communication Engineering, Dehradun, 2015, pp. 535-540
29. Moustafa, Nour, and Jill Slay. "A hybrid feature selection for network intrusion detection systems: Central points." *arXiv preprint arXiv:1707.05505* (2017).
30. Dinil Mon Divakaran, Kar Wai Fok, Ido Nevat, and Vrizlynn L.L. Thing. (2017) "Evidence gathering for network security and forensics ". *Digit. Investig*. 20, S (March 2017), S56-S65.
31. K. Wang, M. Du, Y. Sun, A. Vinel and Y. Zhang (2016)"Attack Detection and Distributed Forensics in Machine-to-Machine Networks," in *IEEE Network*, vol. 30, no. 6, pp. 49-55.
32. Moustaf, N., Jill Slay (2015) "Creating novel features to anomaly network detection using darpa-2009 data set." *Proceedings of the 14th European Conference on Cyber Warfare and Security. Academic Conferences Limited*.
33. Rose, K.; Eldridge, S. & Chapin, L (2015) "The Internet of Things: An Overview".
34. M. M. Hossain, M. Fotouhi and R. Hasan (2015) "Towards an Analysis of Security Issues, Challenges, and Open Problems in the Internet of Things," *2015 IEEE World Congress on Services*, New York City, NY, pp. 21-28.
35. Justin Shattuck, Sara Boddy (2016) "THREAT ANALYSIS REPORT DDoS's Latest Minions: IoT Devices", F5 LABS, Volume1.
36. SCHNEIER, B 2017, 'BOTNETS of Things', MIT Technology Review, 120, 2, pp. 88-91, Business Source Premier, EBSCOhost, viewed 24 August 2017.
37. Eyal Ronen, Colin O'Flynn, Adi Shamir, Achi-Or Weingarten (2016) "IoT Goes Nuclear: Creating a ZigBee Chain Reaction" Cryptology ePrint Archive, Report 2016/1047.
38. M. M. Hossain, M. Fotouhi and R. Hasan (2015) "Towards an Analysis of Security Issues, Challenges, and Open Problems in the Internet of Things," 2015 IEEE World Congress on Services, New York City, NY, pp. 21-28.
39. Yin Minn Pa Pa, Shogo Suzuki, Katsunari Yoshioka, Tsutomu Matsumoto, Takahiro Kasama, and Christian Rossow (2015) "IoTPOT: analysing the rise of IoT compromises." In Proceedings of the 9th USENIX Conference on Offensive Technologies (WOOT'15), Aurélien Francillon and Thomas Ptacek (Eds.). USENIX Association, Berkeley, CA, USA, 9-9.
40. Ashkan Rahimian, Raha Ziarati, Stere Preda, and Mourad Debbabi (2013) "On the Reverse Engineering of the Citadel Botnet." In Revised Selected Papers of the 6th International Symposium on Foundations and Practice of Security - Volume 8352 (FPS 2013) Springer-Verlag, New York, Inc., New York, NY, USA, 408-425.
41. Amir Houmansadr, Nikita Borisov (2013) "BotMosaic: Collaborative network watermark for the detection of IRC-based botnets", Journal of Systems and Software, Volume 86, Issue 3, Pages 707-715, ISSN 0164-1212.
42. "Weka tool", http://www.cs.waikato.ac.nz/ml/weka/, August 2017.
43. Moustafa, N. and Slay, J., 2016. The evaluation of Network Anomaly Detection Systems: Statistical analysis of the UNSW-NB15 data set and the comparison with the KDD99 data set. Information Security Journal: A Global Perspective, 25(1-3), pp.18-31